\documentclass[aps,showpacs,amsmath,amssymb, preprint]{revtex4}
\usepackage{ulem}
\usepackage{amsmath}
\usepackage{graphicx}
\usepackage{color}
\begin{document}
%
\title{Hysteresis Switching Loops in Ag$-$manganite memristive interfaces}

\author{ N.~Ghenzi$^1$, M.~J.~S\'anchez$^2$, F.~Gomez-Marlasca$^1$, P.~Levy$^1$
and  M.~J.~Rozenberg$^{3,4}$ }

\affiliation{$^1$GIA and INN, CAC - CNEA, 1650 - San Mart\'{\i}n, Argentina, \\
$^2$Centro At\'omico Bariloche and Instituto Balseiro, CNEA, 
8400 - San Carlos de Bariloche, Argentina,\\
$^3$Laboratoire de Physique des Solides, UMR8502 
Universit\'e Paris-Sud, Orsay 91405, France,\\
$^4$Departamento de F\'{\i}sica Juan Jos\'e Giambiagi, 
FCEN, Universidad de Buenos Aires,
Ciudad Universitaria Pabell\'on I, (1428) Buenos Aires, Argentina.}

%
%
\begin{abstract}
\textbf{Multilevel resistance states in silver$-$manganite interfaces are studied both 
experimentally and through a realistic model that includes as a main ingredient the  oxygen vacancies diffusion under 
applied electric fields. 
The switching threshold and amplitude studied through Hysteresis Switching Loops are 
found to depend critically on the initial state. The associated  
 vacancy profiles further  unveil the prominent   role of the effective electric field acting at the interfaces.
 While experimental results validate main assumptions of the model, the simulations allow to disentangle
the microscopic mechanisms behind the resistive switching in metal$-$transition metal oxide interfaces.}


\end{abstract}
%
\maketitle
\textit{Introduction}

Manganese based oxides with perovskite structure (manganites) are one of the promising families of transition metal oxides that 
exhibit resistive switching (RS) phenomena \cite{waser,sawa,waser2} \textit{ i.e a reversible and nonvolatile change in the 
resistance after the application of a pulsed electric stimulus (voltage or current)}. 
In addition to two well distinguishable  ``on'' and ``off''  resistance values, multilevel memory states are found. \cite{beck} 
This property, together with demonstrated downsizing possibilities, low power consumption and high 
speed switching, qualify manganite based devices as possible candidates for resistance 
random access memories (RRAM), motivating a huge amount of experimental activity. \cite{baek,liu,beck,watanabe,choi}  
In contrast, few theoretical publications shed light on the microscopic mechanisms behind the RS in 
oxide compounds. \cite{ris,ris2,real,jeong,waser2,nian,tulina,chae}

Recent experimental evidence indicates that the RS effect in metal-manganite devices is due to
the memristive properties of the interface between 
the metallic electrode and the oxide.\cite{hp} Memresistance is a property of a material that amounts to a past-history
dependence of the current magnitude of its resistance. \cite{chua}  
This effect at the interfaces is believed to be driven by the redistribution of oxygen vacancies under the action of  
the applied electric fields. \cite{sawa,szot,seong,nian,real}
In the specific case of conductive manganites, the presence of oxygen vacancies severely disrupts Mn-O links, 
enhancing the resistivity. \cite{nian} A positive 
electric pulse applied at one electrode may cause the migration of oxygen vacancies located in the vicinity of the interface, 
producing a decrease of the contact resistance, due essentially to the recomposition of Mn-O links. \cite{nian} On the other 
hand, negative electric pulses produce vacancies at the interface by repelling oxygen ions.

The Hysteresis Switching Loop (HSL) is a protocol for electrical measurements which was shown to be a powerful tool to 
test the role of oxygen vacancies in RS phenomena.  \cite{nian}
The HSL consists on the measurement of the remnant resistance state (read operation) 
obtained after pulsing (write operation) in a loop sequence 
(i.e., apply increasing  positive pulses  up to a maximum value, after which 
the amplitude is decreased, then the polarity is reversed until a negative maximum is reached,
to finally return back to zero).
The remnant resistance state is measured after every pulse by means of a small bias current.
Here we study such  resistance states at silver $-$manganite interfaces, focusing on their multilevel capabilities.

Our experimental 
results show that the switching threshold and switching resistance amplitude are  both non-trivially determined by the past
resistance history of the interface, suggesting that the local distribution of vacancies near the contact oxide 
interface might 
determine the  main features of the switching response. 
Based on the model proposed in Ref.\onlinecite{real}, that incorporates as a key ingredient the migration of oxygen 
vacancies in a nanoscale 
vicinity of the metal oxide interfaces, we perform numerical simulations that reproduce the experimental data 
remarkably well and 
provide important confirmation to the theoretical assumptions of the model. They indicate that the migration of 
oxygen vacancies
is produced due to the strong electric fields that build up at
the electrode-manganite interface, and is at the origin of the
most significant resistive changes. This fact could provide
valuable guidance to decrease and optimize the strength of
stimulus threshold required for switching in practical devices.

\textit{Experimental Results}

\begin{figure}
\centerline{\includegraphics[width=8cm]{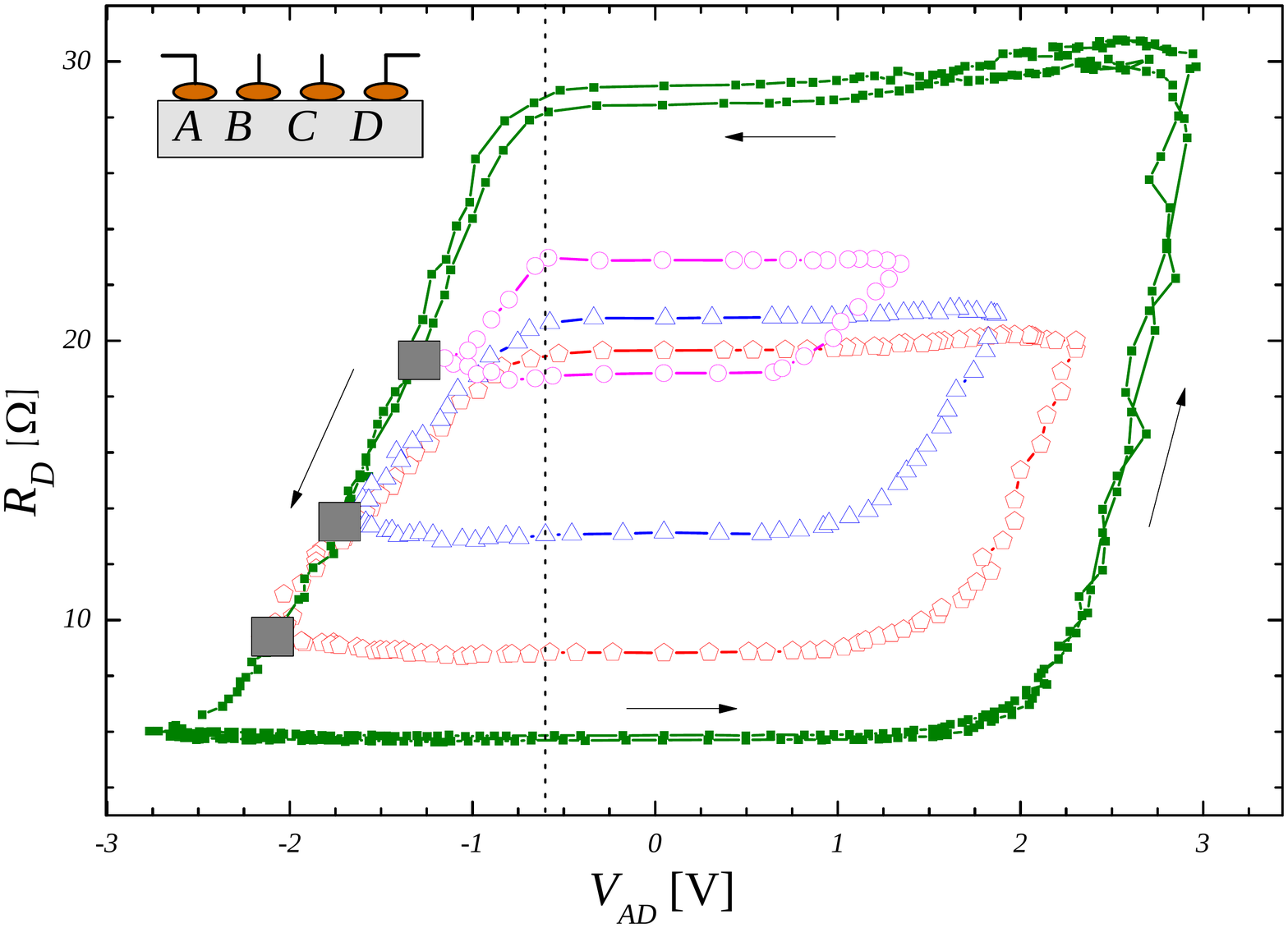}} 
\caption{color online: Resistance $R_D$ at the contact D, measured using a bias current of $I_b=$ 1 mA, as a function of 
the voltage drop  $V_{AD}$ produced  by the electrical pulse previously applied. 
After performing the major loops (filled symbols), minor Hysteresis Switching Loops were obtained starting at selected states 
marked by squares. The vertical dotted line indicates the threshold voltage $V_{th_{-}}$ for switching from H to L.
Inset: Multiterminal electrode configuration used for measuring the interface resistance.} 
\label{fig1}
\end{figure}

We study silver - manganite interfaces by means of a 3 terminal procedure. Millimeter sized contacts were hand painted 
on top of a
bulk  $La_{0.325}Pr_{0.300}Ca_{0.375}MnO_{3}$ polycrystalline sample. A scheme of the electrical contacts is depicted in 
the inset of Fig.\ref{fig1}. Electric pulses 
and bias current are applied through A and D contacts, and voltage measurements are  
independently acquired at the respective terminals. Here we show only results obtained for the D pulsed electrode. 
Measurements at the A pulsed contact are qualitatively similar
and show complementary behavior with respect to contact D \cite{quintero,marlasca}. 
Pulsing was performed with a Keithley 2400 source - meter, while remnant data were 
acquired with an Agilent 34407 data acquisition / switch unit.

Initially, to induce RS on the virgin sample, a set of pulses 
of a given amplitude and polarity is applied between electrodes A and D,
followed by a similar set with the opposite polarity \cite{quintero,marlasca}, 
and repeating the procedure several times.
After this, pulsing in a loop mode (i.e. the HSL) was the electrical protocol used to switch the interface 
resistance \cite{nian,marlasca} (write operation). Each squared
pulse of 10 ms time width was followed by a small bias $I_b$ applied some 10 seconds after the pulse to 
obtain the remnant resistance curve 
(read operation). Heating effects appear to be negligible, as determined from dynamic measurements.

Figure \ref{fig1} depicts the  resistance values $R_{D}=V_{CD}/I_{b}$ for the
interface at electrode D as a function
of the pulsing voltage strength applied between electrodes A and D.
$I_{b}$ is the small bias current applied after each pulsing.
We begin the pulsing protocol by performing a major loop that
corresponds to the maximal pulsed-voltage excursion.
 As observed, the resistance  $R_{D}$ exhibits two well defined states
(around 30  $\Omega$ and 6 $\Omega$), namely the high H and low L resistance  states, 
and two rapid transitions through a multitude of
stable intermediate states.

Starting at the L state, we observe a certain
threshold voltage of positive pulsing
stimulus, $V_{th_{+}} \approx$ 2 V, that is required
to initiate the rapid upward change in the resistance.
The total resistive switching produces a factor - of - five change in $R_{D}$.
Once the H state has been achieved at the maximal
positive pulsing strength, the loop mode tests the stability of the
state against pulses of gradually decreasing amplitude and then
its coercivity, i.e. the stability against pulses of opposite polarity with increasing strength.

We observed negligible changes in $R_{D}$ while the polarity of the stimulus remains positive.
More significantly, the interface resistance also shows a significant coercivity, as the
transition from H to L only begins when a  small but non-zero
negative pulsing $V_{th_{-}}$ is applied (indicated by a vertical dashed line in
Fig.\ref{fig1}).

During the rapid transition, the interface traverses a multitude of different resistance
states which are all (meta)stable as we shall see later on. This feature is a clear
signature of the memristive properties of the interface. 

Eventually the L resistance value
$R_{D} = 6   \Omega $ is reached at the maximum negative polarity pulsing, and
it remains stable through the rest of the loop. This is, both, when decreasing
negative pulses towards zero, but also upon  changing to positive polarity and
increasing the pulsing strength up to $V_{th_{+}}$. Again, this fact 
 demonstrates the significant coercivity of the resistance state.
 Interestingly, the
values of $V_{th_{+}}$ and $V_{th_{-}}$ have different magnitudes and
also, as we shall see below, different dependence on the applied voltages.
This is not surprising, due to the directional asymmetry of the ionic migration 
at a each interfacial region. 

Upon repeating the HSL protocol, a second major loop is formed almost perfectly on top of the
previous one, demonstrating a good reproducibility and control of the memristive effect
at the interfaces of our device.

To further characterize the intermediate multilevel
states during the H to L transition, 
 additional measurements were performed sequentially 
ranging the pulsing stimulus between lower voltage
limits (i.e. "minor loops"). Different initial states on the
major loop were used for starting each of the minor loops.
After measuring a set of three minor loops for a given
voltage excursion, always a set of three new major loops were performed
to reset the system. All loops of the same set (minors and major) 
showed similar reproducibility
characteristics as the one already shown for the first major loop. Therefore,
for the sake of clarity we only show in  Fig.\ref{fig1} a representative minor loop
for each set.

As observed, upon performing minor loops, new sets of intermediate stable L and H resistance states are obtained. Note that these 
new L states have the same $R_{D}$ value of the initial intermediate state. The corresponding H value is found to be always lower 
than that of the major loop. Additional aspects of these sequentially obtained data are to be pointed out:
\begin{enumerate}
\item
The stability of the multilevel intermediate resistance
states is clearly established.
Minor loops have the same qualitative shape as the major one, their actual 
 H/L (or ON/OFF)  resistance ratio being proportional to the sweep range of the stimulus. 
In addition, the reversible character of the complete sequence of a major loop followed by a minor one is remarkable.

\item
For positive stimulus, the threshold for switching $V_{th_{+}}$ is
higher when the initial state resistance $R_{D}$ is lower. Note that
this resistance value becomes the L state of the corresponding
minor loop.
\item
However, for negative stimulus the situation is different. The dotted line on Fig.\ref{fig1} indicates that upon reversing the 
pulsing polarity, the remnant resistance
starts decreasing at a voltage value $V_{th_{-}}$ that is almost
independent of the initial resistance level.

\end{enumerate}

We now turn to 
the comparison of these experimental results 
with simulations performed using the model introduced in Ref.\onlinecite{real}.

\textit{Model Simulations}

We here describe the main ingredients that built up  the theoretical model  together  with 
the basic equations employed in the numerical simulations. For  further details we refer the reader to 
Ref.\onlinecite{real}.

The model mimics the active region for conduction as a one dimensional resistive network, or chain, 
of $N$ links, 
each one characterized by a certain concentration of oxygen vacancies. 
This 1d model may be considered as a
drastic simplification of the actual 3d geometry. However, this assumption is
supported by the fact that experimental evidence indicate that the conduction in the low
resistance state is highly inhomogeneous, and takes place
along preferential and directional paths \cite{sawa}.

The first and last $N_I<< N$ links define  the left (L) and right (R) interfacial regions between 
each metallic electrode  and the bulk material (B) that is described by $N - 2N_I$ links.    

The main idea behind the model is that the local oxygen vacancy concentration
of a domain, $\delta_i$, determines the
resistance  of the respective network link $\rho_i=  A_\alpha \delta_i $.
The constants $A_\alpha$ relate vacancy concentration and resistivity, and
depend on the material and on the region. In principle the could be computed
from a microscopic approach that incorporates geometrical and 
material specific information. This important but technically challenging issue 
is beyond the scope of the present phenomenological modeling.
The constants are defined as $\alpha = B$ if $i$ belongs to the bulk ($N_I < i \leq N-N_I$), 
$\alpha = L$ if $i$ is in the left interface ($i \leq N_I$)
and $\alpha = R$ if  $i$ is in the right interface ($N-N_I < i \leq N$). 
The parameters $A_\alpha$  are taken to be  $A_{L,R} >> A_{B}$   to 
stress that both interfaces are much more resistive  than  the bulk region. 
Indeed, experiments indicate  that
a Schottky  type of barrier  formed between a metallic electrode and the oxide 
could be the origin of the contact resistance \cite{fujii,sawa2,chen}. 
The total two terminal resistance is defined as  $ R_T = c \sum_{i=1}^{N} {\rho_i} = 
c \sum_\alpha \sum_{i\in \alpha}{A_\alpha \delta_i} \equiv R_{L} + R_{B} + R_{R}$, 
where $c$ is a geometrical factor that we take equal to unity without loss of generality.

Under an external  applied stimulus (typical experimental voltage  or current protocols),
 the  voltage drops at the network domains  create local electric fields
that promote the diffusive motion of the positive charged oxygen vacancies. 
In order to specify the model equations we shall consider  as the  stimulus
a given voltage protocol $V(t)$ applied between L and R electrodes  (see below).

The local dynamics for the diffusion of
vacancies  is ruled by the following equation,
\begin{equation}
p_{ij} = \delta_i (1-\delta_j) \exp(-V_0 + \Delta V_{i}) \;,
\label{proba}
\end{equation}
that specifies the probability for  transfer of 
vacancies from domain $i$ to a nearest neighbor domain $j=i \pm 1$.

The probability Eq.\ref{proba}, is proportional to the concentration of vacancies
present in domain $i$, and to the concentration of ``available vacancy
sites'' at the neighbor domain. 
In addition, Eq.\ref{proba} is also proportional to the Arrhenius factor $\exp(-V_0 + \Delta V_{i})$, where $V_0$ is a 
dimensionless constant related to the activation energy for vacancies diffusion and  $\Delta V_i$ is the 
local potential drop ${\Delta V}_i (t) = V_{i+1}(t) - V_i(t)$ with $V_i(t) = V(t) \rho_{i} /R_{T}$.

Starting from an initial vacancy concentration profile $\delta_i (0)$, the numerical simulations are performed through the
following steps: (i) at each simulation time step $t$ a given external voltage $V(t)$ is applied between the electrodes,
that are assumed as perfect conductors,
(ii) we compute the local voltage profile $V_i(t)$ and the voltages drops 
${\Delta V}_i (t)$.
(iii) we use Eq.~(\ref{proba}) to compute all the oxygen vacancy transfers between nearest neighboring 
domains, and update the values $\delta_i(t)$ to a new set of concentrations $\delta_i(t+1)$.
(iv) we use these new values to recompute at time $t+1$ the local resistivities  $\rho_i(t+1)$  and 
the local voltage drops under the applied voltage
$V(t+1)$, as indicated in the first step. 

A crucial  outcome  of the model \cite{real} is that the local electric fields  created in the vicinity of 
the interfacial regions are much more intense than in the bulk. Thus as we shall show below, the motion of vacancies 
 is enhanced in the  vicinity of the interfaces, producing dramatic changes in its local resistivities. 
As a consequence, the change in the two terminal resistance, $ R_{T}$, are mainly 
due  to  variations of the interfacial resistances, $R_{L} + R_{R}$.
This is in perfect agreement with  the experimental findings that suggest  the interfaces as the active regions for 
resistive switching.  \cite{nian, quintero, ignatiev,chen}

To compare  results  obtained by model simulations with  the experimental ones already described, we shall concentrate 
 on the right interfacial resistance $R_{R}$, that by analogy 
 corresponds to  the interface in contact to the  D electrode 
in the experimental set up. Following Ref.\onlinecite{real}, in our simulations we take the total number  of sites 
$N =100$ where the first and last  $N_{I}= 10$ sites define the   L and R interfaces respectively.
 We analyze a symmetric configuration,  assuming  identical  interfaces for 
both (Ag) electrodes,  setting 
$A_R = A_L = 1000 >> A_B =1$.  In addition  we choose $V_{0}=16$,
to provide a non-negligible but slow diffusive contribution 
to the evolution of $\delta_i$. 

\textit{Comparison with experimental data}

\begin{figure}
\centerline{\includegraphics[width=8cm]{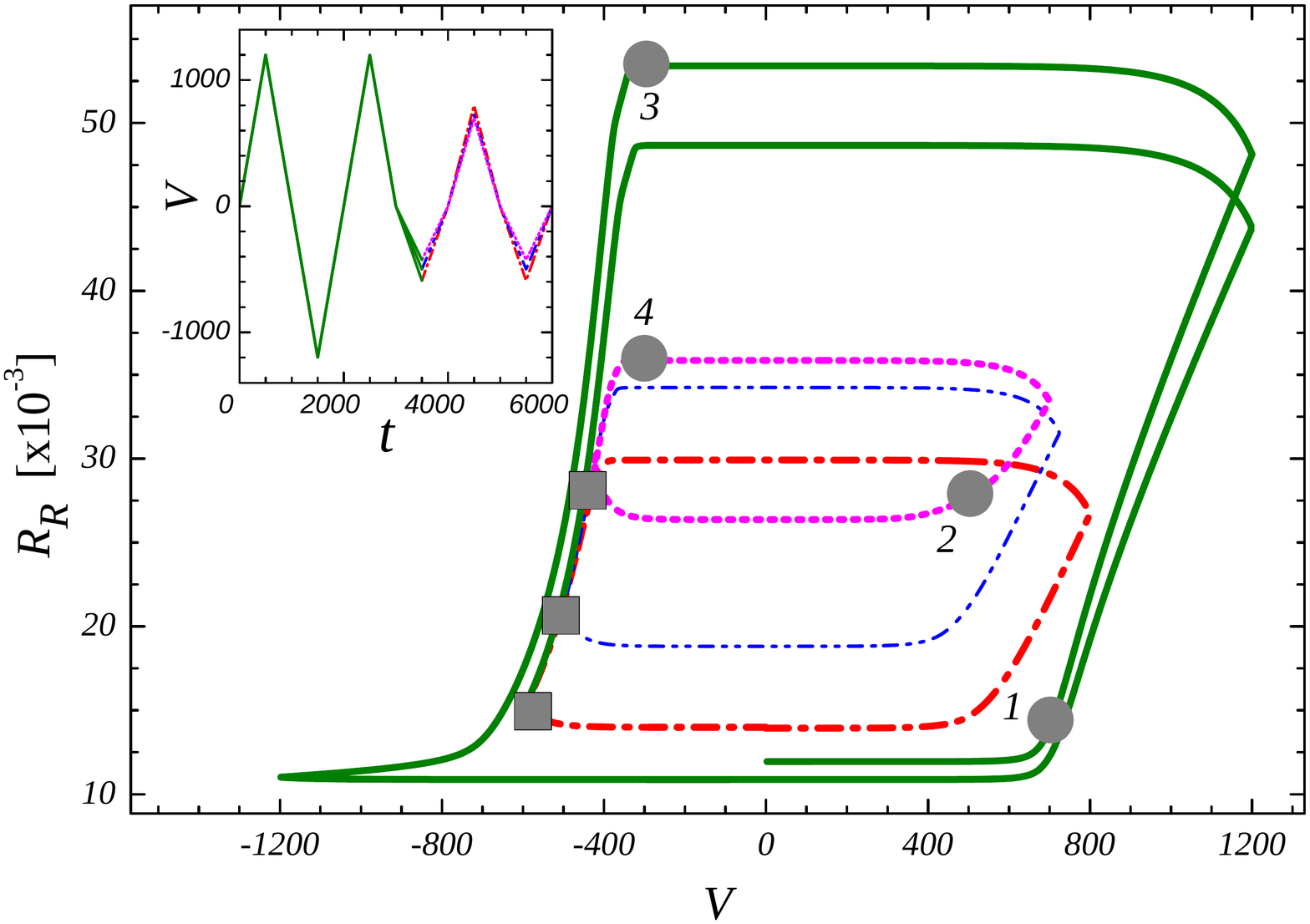}} 
\caption{Resistance $R_R$ at the right  interface as a function of the stimulus $V(t)$, as
obtained from model simulations. Squares depict initial states for minor loops. Numbered dots indicate threshold voltages for L to H 
and H to L transitions  in selected HSL (see text for details). Similarly to the 
experimental case, the threshold voltages for switching from L to H depend on the initial state.
Inset: Voltage protocol sequences (major loop followed by a minor loop) as a function of time steps.} 
\label{fig2}
\end{figure}

Figure \ref{fig2} shows  in a solid line a  HSL  of the  resistance $R_{R}$  
as a function of applied voltage protocol $V(t)$. Each   cycle  is defined by a  linear ramp 
$0 \to +V_{\rm max}\to 0 \to -V_{\rm max} \to 0$  of $\tau = 6000$ time steps of duration
and $V_{\rm max}=1200$, to supply a sufficiently large electric stimulus. \cite{note} 

The agreement  with the experimental HSL obtained in Fig.\ref{fig1}  is evident. 
As the voltage is increased from $V= 0$, and after 
surpassing the threshold value (marked by  dot $\#$ 1 
in Fig.\ref{fig2}) the high resistance state H is stable and is attained 
at V$_{max}$, while multiple states are traversed in between, i.e. during the transition 
from L to H. 
The H state remains stable when
decreasing the voltage from V$_{max} \to 0$.
Significantly, it also remains stable after the inversion
of polarity and up to a small but finite threshold $V_{th_{-}}$
that is indicated with the dot $\#$3. Similarly to the experimental
case described before, beyond $V_{th_{-}}$ the transition from H to L
state starts and the interface goes through a multitude of
stable states of decreasing resistance.
At the maximum negative voltage, the interface reaches the L state
of lowest resistance. Then, it remains stable as the voltage is ramped
up from $-$V$_{max} \to 0$, and also after the voltage polarity inversion
up to the $V_{th_{+}}$. Notice, however, that the new loop does not go
precisely on top of the previous one.
Thus upon starting  a new cycle for the  
same voltage sequence, a small offset  is obtained in the resistance values. We shall come back to this point later
 when discussing
the  evolution of the vacancies profiles.

Simulations of minor loops response were also performed  ranging the voltage profile
in a sequence $ -V_{m1}\to 0 \to V_{m2} \to 0  \to  -V_{m1}$. 
The different initial values indicated by the squares in Fig.\ref{fig2} , 
where chosen along  the  major loop in order to mimic the experimental protocol. Each   starting point 
  has   associated a particular vacancy profile $\delta_{i}^{s}$  that  works as the initial
state   for the numerical simulations that produce each minor HSL, identified  in Fig.\ref{fig2} with different traces.

Interestingly, several qualitative features of intermediate states  
in the simulated HSL reproduce nicely the experimental findings. 
The value of the predicted voltage threshold $V_{th_{+}}$ for L to H transition
in the model decreases for higher initial
resistance values (i.e.
higher resistance of the L state). In addition, we also observe that
the threshold voltage $V_{th_{-}}$ for
the H to L transition is almost independent of the H
resistance value. These two features are in good qualitative agreement with our experimental findings
described above. 

Finally, it is interesting to correlate this switching behavior with the respective evolutions
of vacancy profiles and 
local electric fields that develop along the conductive path.

\begin{figure}
\centerline{\includegraphics[width=8.5cm]{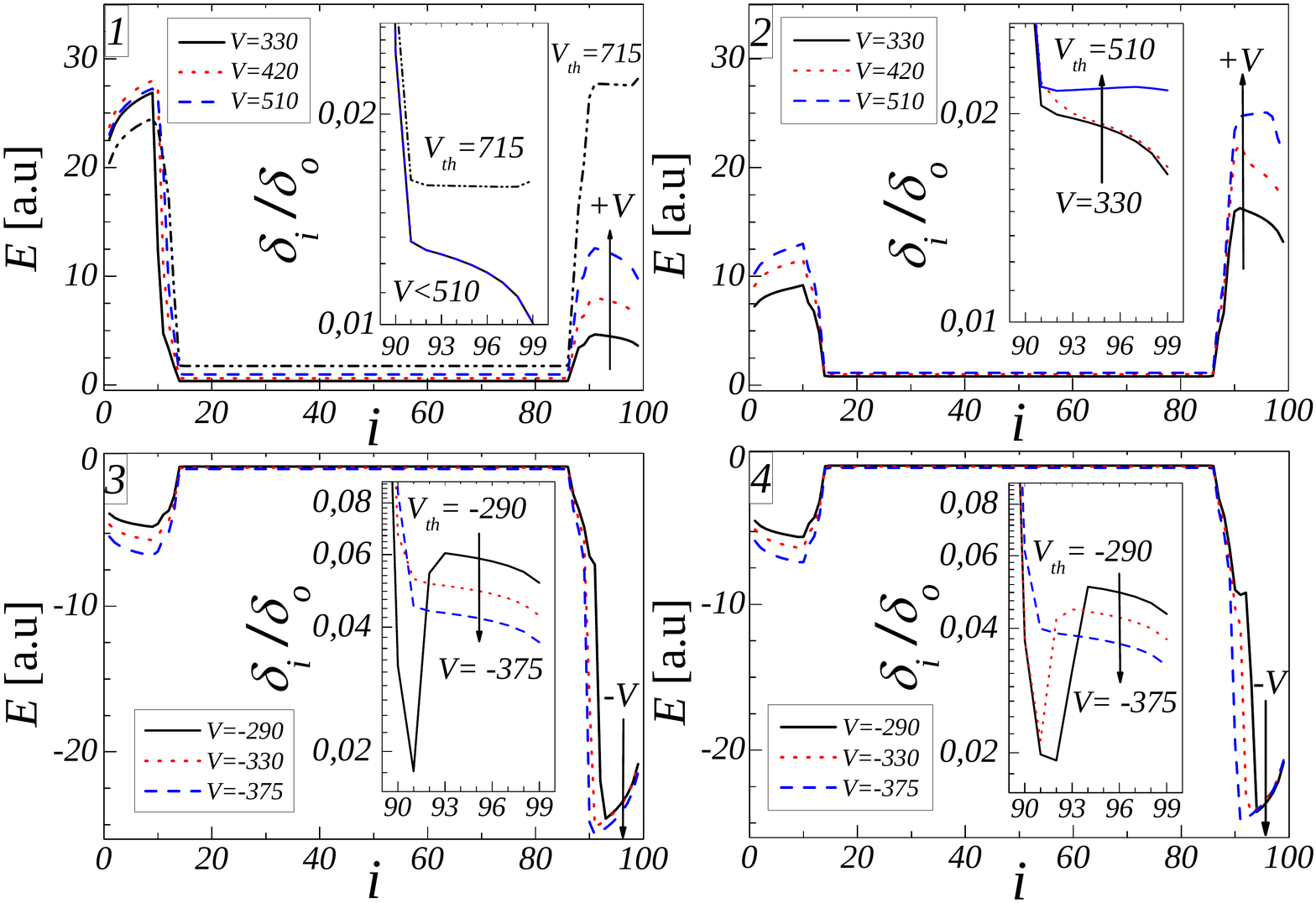}} 
\caption{color online: Panel $i=1,2,3,4$: Electric field profiles $E_{i}$ along the sample for  different values of $V$, close
 to the threshold voltages  labeled  in Fig.\ref{fig2} as $\#$
1,2,3 and 4, respectively. The inset in each panel shows the vacancy concentration 
profiles (in units of  $\delta_i (0)= 10 ^{-4}$) along the R interface
for the selected values of $V$.} 
\label{fig3}
\end{figure}

Figure \ref{fig3} depicts the electric field profile  $E_{i} \propto \rho_i V/R_{T} $ along the sample, 
where $V$ is the applied voltage and $R_T$ is the total two-terminal resistance. The data of the figure
corresponds to states that are
close to  the  thresholds values for the L to H (panels 1) and 2)) and H to L  (panels 3) and 4)) transitions 
for the selected major and minor loops (panel numbers correlate with the points indicated in Fig.\ref{fig2}). 
In coincidence with   previous studies \cite{real}, 
Fig.\ref{fig3} shows that the  interfaces are  the regions in which 
locally enhanced electric fields built up and  promote the motion  of vacancies.
On the other hand,  due to the metallic  character of the bulk and its lower resistivities
($ A_{B} << A_{R} = A_L $),  negligible electric fields act  well inside
the bulk region.
However, the accumulation of vacancies near the transition regions L-B and R-B
lead to significant field in the first few links that enter the bulk \cite{real}. This feature can be clearly seen in the
insets of each panel of Fig.\ref{fig3}.
The continuity and penetration of the field E across these internal interfaces
allows for migration of vacancies to and from the bulk reservoir.

We shall concentrate  in  the vacancy profiles  along the 
R interface, but a similar analysis could be done for the left interface.
In the respective insets of Fig.\ref{fig3}, we plot the vacancy concentration profiles 
(normalized to the constant initial value $\delta_{i}(0)$) along the 
R interface (sites $i=$ 90 to 100).

Below we shall  analyze  how  vacancies and electric field profiles evolve close to the 
L(H) to H(L)  HSL transition.
 
In  panels 1) and 2) of Fig.\ref{fig3} we compare the vacancies and electric field profiles 
for values of $V$ close to the threshold $V_{th_{+}}\sim 715$ and $\sim 510$ (dot $\#$ 1 and 2 
respectively in Fig.\ref{fig2}) for the L to H transitions. 
We observe that at  the respective threshold voltages, the vacancies density  profiles of the
relevant R interfacial region $\delta_i$
saturate in both cases to a constant value. This is the state reached before vacancies start to
migrate out of the resistive interface into the much less resistive bulk.
Notice that the constant value is higher in (2) than in (1), corresponding to the
higher resistance value of the L state in the minor loop respect to the major one.

Comparing the local electric field   at the R interface,
we find that  {\bf a similar  value}  $E_{th} \sim 22 -27$ a.u. is attained at both  L to H transitions. 
In addition we have also  checked that the same  values of $E_{th}$ are obtained
for other minor HSL at the  respective L to H transitions. Taking into account that in our simulations
we have taken  $V_{0}=16$ as the   activation energy for diffusion, this value of $E_{th}$ 
provides  a significant  migration  of vacancies into  the R interface in order to produce the L to H transition 
in the different HSL.

Note in addition that, for the different values
of applied $V's$, the local  electric field at the R interface exhibits important relative variations  in comparison 
to  the left interface. This shows that the switching process occurs rather independently at
each interface and that the switching occurs fast when the threshold level is reached.

Although at both threshold voltages
the local electric fields at the R interface have rather similar values,
at the left interface, $E$ in panel 1) almost duplicates the values in panel 2). This accounts for a
higher  value of the threshold voltage in panel 1) as compared  to   panel 2).
 
Next, we focus on the H to L transition. Vacancy and electric field 
profiles for the states labeled by dots $\#$ 3 and 4 in 
Fig.\ref{fig2} are depicted respectively in panels 3) and 4) of Fig.\ref{fig3}, for  
three values of $V$ close to the threshold voltage 
$V_{th_{-}} \sim -290$ for the H to L transition in both HSL.
Electric field profiles look qualitatively  similar for all $V$'s investigated.  
The large  (and negative) values of $E$  along the R interface are a signature 
of the H resistance state. In addition due to  the similar  electric field profiles along the whole sample 
the same  threshold value ($V_{th_{-}} \sim -290$) is obtained for  both   H to L transition in panels 3) and 4), 
in remarkable agreement with the experimental data. 

Vacancy profiles shown in the insets of  panels 3) and 4) of Fig.\ref{fig3} are  also qualitatively similar, although 
the profile in  3) exhibits a higher and broader distribution than in 4), consistent with a higher resistance value. 

Note that the vacancy profiles for  both L to H transitions (panels 1) and 2)) are quite  smooth, 
while those for the  H to L transitions (panels 3) and 4)) are sharper and non monotonic. 
Thus the  H resistance state in each HSL  has associated 
a rather complex vacancy distribution as a consequence of the  enhanced  internal electric field at the interface.  

Upon cycling periodically, a small concentration of vacancies  that migrate from the interfaces towards the bulk remain 
confined at the bulk due to relatively small electric fields acting at the bulk.
Once there, they no longer contribute to interfacial switching. 
This explains the small offset in the resistance values
obtained, experimentally and in the simulations, in the HSL after  
each cycle is completed  (see Figs.\ref{fig1} and \ref{fig2}).
This effect is possibly responsible for the long time degradation of the 
switching ON/OFF ratio. A systematic study of this feature is therefore
important and will be the focus of future work.

Recently, Das et al. studied  the threshold stimulus dependence for 
switching in manganite thin films 
with Ag contacts by injecting pulses of opposite polarity. \cite{das} 
They obtain a sample to sample variation, and they emphasize 
the sigmoid like (universal) 
character of the dependence of the RS behavior with 
the applied stimulus \cite{das}. In addition,
 our results suggest that the choice of the initial state, i.e.
the resistance value within the H to L transition branch of the major loop,
is a key parameter in order to determine both, threshold stimulus $V_{th_{+}}$ value, 
and the resistive ON/OFF ratio. 
We have also demonstrated a full control of these features as we were
always able to restore back the system from any minor loop to the major one.

\textit{Final remarks}

Multilevel  resistance states in pulsed Ag- manganite contacts have been directly related to different oxygen 
vacancy profiles and local electric fields  at the interface. 
The strong dependence of the threshold stimulus $V_{th_{+}}$ on the initial state is a signature of the role played by 
detrapping or electric field enhanced migration of oxygen vacancies in the underlying RS mechanism. 

As a summary, we list main findings of this work:

- stable multilevel states exhibiting RS are obtained and controlled experimentally following a loop pulsing procedure; 

- a remarkable qualitative agreement between experimental and simulated data has been obtained, 
providing further validation to the model introduced in Ref.\onlinecite{real}.  

- control of RS switching characteristics such as ON/OFF ratio and coercivity were demonstrated.

Our results suggest that different initial vacancy configurations could be tailored by the application
of other voltage protocols than the one used here for the major loop. Thus one may conceive that an optimal
low threshold stimulus for a given desired ON/OFF switching ratio could be obtained. 
Moreover, the model validation shown here opens the way for a rapid testing of other initial vacancy 
configurations, suggesting alternative electro forming-like procedures.  
We believe that our work demonstrates the usefulness of theoretical modeling
of the resistive switching phenomena as a valuable aid to provide
guidance in the
analysis of the experimental results and, eventually, for the design
and optimization  of memory devices.

\begin{acknowledgments}
Support from CONICET (grants PIP 5254/05 and PIP 112-200801-00047) and ANCTyP 
(grants PICT 483/06 and PICT 837/07) is gratefully acknowledged.
We thank  R. Weht for his contribution in the early stage of this work. MJS, PL and MJR are members of CONICET.
\end{acknowledgments}

\begin{thebibliography}{10}
\bibitem{note} In the numerical simulations the calculated values
of resistance at each simulation time step corresponds to the
experimental resistance that is measured with the small bias current.
\end{thebibliography}
\end{document}